\documentclass[elsart12,epsf,eqsecnum]{revtex4}
\begin{document}


\voffset1.5cm


\title{In pursuit of Pomeron loops: the JIMWLK equation and the Wess-Zumino term.}
\author{Alex Kovner and  Michael Lublinsky}

\address{Physics Department, University of Connecticut, 2152 Hillside
Road, Storrs, CT 06269-3046, USA}
\date{\today}

\begin{abstract}
We derive corrections to the JIMWLK equation in the regime where the charge density in the hadronic wave function is small. 
We show that the framework of the JIMWLK equation has to be significantly modified at small densities in order to properly account for the noncommutativity of the charge density operators.
In particular the weight function for the calculation of averages can not be real, but is shown to contain the Wess-Zumino term. 
The corrections to the kernel of the JIMWLK evolution which are leading at small density are resummed into a path ordered exponential of the functional derivative with respect to the charge density operator, thus hinting at intriguing duality between the high and the low density regimes.
\end{abstract}
\maketitle
\section{Introduction.}

The recent years have seen vigorous activity in an attempt to understand high energy scattering in QCD from first principles. The precursors
of this recent wave are the classic papers \cite{glr} and \cite{bfkl} as well as later work \cite{al} and \cite{mv}.

The equation that governs the evolution of the $S$-matrix of a small projectile with energy was derived by Balitsky\cite{balitsky}, and its mean field version by Kovchegov\cite{kovchegov}. 
This equation takes into account perturbative evolution of the projectile wave function supplemented by multiple scatterings of the projectile partons on the target. 
A lot of numerical and analytic work in recent years has been carried out to understand the properties of the solutions of this equation both in the ultraviolet and infrared. The dependence of the saturation momentum on rapidity \cite{ll}, the property of geometric scaling\cite{iancu}, the 
power growth of the total cross section \cite{kw} and the disappearance of the Cronin effect with energy \cite{cronin} are some stark examples of the physical information obtained from this equation so far.

A complementary approach to the high energy scattering problem was initiated in \cite{JIMWLK,cgc}.
It yields the nonlinear evolution equation  (JIMWLK equation) that governs the change of the correlation functions of the color charge density in the hadron wave function. The parameter of the evolution, as before is the rapidity of the hadron. In this approach it is assumed that the charge density in the wave function is high. It has been shown in the last reference in \cite{JIMWLK} and in \cite{cgc}(see also  \cite{mueller}) that when understood as the equation for the wave function of the large target, for processes when this target is probed by a small perturbative projectile, the JIMWLK equation is equivalent to that of Balitsky as far as the calculation of $S$-matrix is concerned.

One of the original motivations to consider the  high energy scattering, is the question how does the scattering amplitude approach the unitarity limit. 
The Balitsky-Kovchegov and JIMWLK equations indeed lead to unitary amplitude: at large energies the $S$-matrix approaches zero (not to be confused with the unitarity of the total cross section in the sense of the Froissart bound, which is not achieved in this framework \cite{kw}). However it is clear that the way the amplitude approaches the unitarity limit is not properly described by these equations. In the framework of the Balitsky-Kovchegov the reason is that the evolution of the projectile wave function is linear, that is gluons are emitted independently by the "valence" partons in the projectile. The unitarization of the amplitude is then achieved only due to the multiple scatterings on the target. However it is clear that when evolved to large enough rapidity, the wave function of the projectile becomes dense, and at that point the gluons have to be emitted coherently from the charge density distribution in the target, rather than independently from every parton. At high density the gluons are produced less efficiently due to these wave function saturation effects, and this furnishes an additional mechanism for the unitarization of the scattering amplitude over and above multiple scatterings. When the target is large and the projectile is small, the multiple scattering mechanism is more efficient. However in this situation one already at low energy starts from the almost vanishing scattering matrix (black target), and so can not sensibly  study the approach to the black limit. On the other hand, for a small target, the unitarity corrections due to the wave function saturation effects in the projectile become important at the same rapidity as those due to multiple scatterings, and the approach to the
unitarity limit is described incorrectly when they are omitted.

In the framework of the JIMWLK equation one is faced with a similar problem. The equation is derived in the limit of high density. Although its low density limit reduces to the linear BFKL equation, there is no reason to believe that it describes correctly the evolution in the intermediate regime, where the charge density is neither small nor large.
Therefore one cannot consistently use the JIMWLK equation to follow the evolution of the wave function of a "dilute" system all the way until it becomes "dense".

The main physical effects that take place in the intermediate density regime are due to the so-called Pomeron loops. Those are the processes where the gluons, which are emitted in the wave function at an earlier stage of the evolution, subsequently "drop out" of the evolution as their color is "bleached" by other gluons and they cease their ephemeral existense.
The JIMWLK equation contains only part of the Pomeron loops. Although both, the gluon emission and the gluon disappearance processes have their place in the JIMWLK evolution, they are not described properly at small and intermediate densities (rapidities). Proper inclusion of these processes must also restore the $t$-channel unitarity which is not preserved by the JIMWLk equation \cite{shoshi1}. One would like to have a better handle on the Pomeron loops, since they are clearly important for the evolution of the scattering amplitude. 

To that end one has to understand the evolution of the hadronic wave function at arbitrary color charge density. This is a very challenging problem which at present is not solved.
Recently a first step in this direction was undertaken\cite{iancu1},\cite{shoshi},\cite{misha}. These works derive a correction to the JIMWLK equation, which can be described as the first term in expansion in powers of the functional derivative with respect to the color charge density.
In the present paper we go beyond this first step and resum an infinite number of such terms.  
The terms we find do not give the complete kernel of the JIMWLK equation at arbitrary density. We do however resum all terms which are leading in the low density regime (plus a little bit extra - see below). Along the way we find that the framework of the JIMWLK equation has to be modified to accomodate the low density regime. In particular, since one cannot treat the charge density operators as commuting when the density is not large, the quantum averages can not be written simply as weighted averages, (functional integrals) over the real functional $W[\rho(x)]$. Rather the charge density variable in the functional integral must be considered as dependent on an extra variable $t$ and the weight functional $W$ must contain a phase which is given by the Wess-Zumino term. We also show  that this auxiliary variable $t$ plays exactly the same role as the longitudinal coordinate $x^-$ in \cite{jkmw} and \cite{cgc}, namely it resolves the ambiguity associated with the ordering of quantum fields. Thus for all intents and purposes $t$ can be identified with $x^-$.

The correction terms we derive are resummed in a compact and suggestive expression. It has a very interesting structure which hints at an intriguing duality between the low and high density regime. 

This paper is organized as follows. In section 2 we derive the correction to the correlation functions of the charge density operator in terms of averages of quantum operators. In section 3 we translate this into the corrections to the functional evolution equation for the functional $W[\rho]$. We also explain how the extra variable $t$ and the Wess-Zumino term arise in this context. Finally in section 4 we discuss several issues and questions which are prompted by our results.

\section{High energy hadronic wave function and the evolution of the correlators of color charge density.}

The JIMWLK evolution equation is a functional evolution equation for the weight functional $W[\rho^a(x)]$. It was shown in \cite{cgc} that the  form of the evolution is somewhat simpler when $W$ is considered to be a function of the phase of the scattering matrix $\alpha$ rather than $\rho$. For our purposes in this paper however we find it more convenient and straightforward to use the original formulation in terms of $\rho$\cite{JIMWLK}. The functional $W$ has the meaning of the probability density to find a given configuration of the charge density,  so that expectation value of any observable
$O[\rho]$
in the hadronic wave function is given by
\begin{equation}
\langle O\rangle=\int D\rho^aO[\rho(x)]W[\rho(x)],\ \ \ \ \ \ \ \int D\rho^a(x)W[\rho(x)]=1
\label{average}
\end{equation}
Here $x_i$, $i=1,2$ are the transverse coordinates, and $\rho^a(x)$ is the surface charge density, defined as the integral of the three dimensional charge density over the longitudinal extent of the hadron
\begin{equation}
\rho^a(x)=\int dx^-\rho^a(x,x^-)
\end{equation}
The JIMWLK evolution equation reads
\begin{equation}
{\partial\over\partial Y}W[\rho]=\alpha_s\left\{\int d^2xd^2y\chi^{ab}[\rho]{\delta\over\rho^a(x)}{\delta\over\delta\rho^b(y)}+\int d^2x\sigma^a(x){\delta\over\rho^a(x)}\right\}W[\rho]
\end{equation}
The explit form of the functionals $\chi$ and $\sigma$ is not important for our present purposes, except for the fact that they both have expansion in powers of $\rho$, so that at small $\rho$ one has $\chi=O(\rho^2)$ and $\sigma=O(\rho)$.

The equation was derived for parametrically large densities $\rho\propto {1\over \alpha_s}$.
We will now extend the derivation to include also the small density region $\rho\propto 1$. The easiest way to do this is to consider directly the evolution of the hadronic wave function. Although the original derivation in \cite{JIMWLK} was given in the language of the path integral, it can be formulated directly in terms of the wave function \cite{inprep}. One starts at the initial rapidity with a wave function $|\Psi(Y)\rangle=|v\rangle$ which contains only "valence" degrees of freedom, namely those with the longitudinal momentum $k^+$ above some cutoff value $\Lambda$.
This wave function defines the charge density correlation functions
\begin{equation}
\langle v|\rho^{a_1}(x_1)...\rho^{a_n}(x_n)|v\rangle
\end{equation}

When boosted by a small amount, the valence wave function gets dressed by a cloud of the Weiszacker-Williams gluons. 
The change of the wave function can be calculated explicitly. The evolved wave function has the following structure\cite{inprep}
\begin{equation}
 |\Psi(Y+\delta Y)\rangle=\left\{\left[1-{\delta Y\over 2 \pi}\int d^2x(b_i^a(x)b_i^a(x))\right]+i\int d^2xb_i^a(x)\int_{(1-\delta Y)\Lambda}^{\Lambda}{dk^+\over \pi^{1/2}| k^+|^{1/2}} a^{\dagger a}_i(k^+, x)\right\}B(a^\dagger,  a)|v\rangle
 \label{wf}
\end{equation}
Here the creation operators $a^\dagger(k^+)$ create gluons with soft momenta, which are not present in the valence state $|v\rangle>$. The field $b$ depends only on the valence degrees of freedom. It is determined as the solution of the "classical" equation of motion 
\begin{eqnarray}
&&\partial_ib_i^a+g\epsilon^{abc}b^b_i(x)b^c_i(x)=g\rho^a_v(x)\nonumber\\
&&\epsilon_{ij}[\partial_ib^a_j-\partial_jb^a_i+g\epsilon^{abc}b^b_ib^c_j]=0
\label{b}
\end{eqnarray}
This is precisely the "classical background field" that appears in \cite{JIMWLK}. The only subtlety is that the commutator in the first equation was dropped in \cite{JIMWLK}, which is indeed appropriate for large charge densities. In this case it is straightforward to see that the commutator term is $O(\alpha_s)$ correction to the first term in the equation and can therefore be nerglected. In fact for small charge densities, $\rho=O(1)$ the field $b$ is of order $g$, and so the commutator term is again negligible. We keep it here for completeness, as 
it can in principle be important in the intermediate regime. The $\rho_v$ in the right hand side of eq.(\ref{b}) is the color charge density of the valence degrees of freedom only, that is of the gluons with longitudinal momentum above the cutoff $\Lambda$. We stress again, that in the present paper we consider $\rho_v$ and $b_i$ to be fully quantum operators which act on the Hilbert space of the valence degrees of freedom.

The term in the curly brackets in eq.(\ref{wf}) is simply an expansion to first order in $\delta Y$ of the coherent operator 
\begin{equation}
C=\exp \left\{i\int d^2xb_i^a(x)\int_{(1-\delta Y)\Lambda}^{\Lambda}{dk^+\over  \pi^{1/2} |k^+|^{1/2}} [a^{\dagger a}_i(k^+, x)+a^a_i(k^+, x)]\right\}
\end{equation}
which affects the shift of the soft modes of the gluon field by the "classical field" $b^a_i(x)$
\begin{equation}
C^\dagger A^a_i(k^+, x)C=A^a_i(k^+, x)+{i\over k^+}b_i^a(x)
\end{equation}
In eq.(\ref{wf}) we have only kept $C$ to first order in $\delta Y$, since we mean to calculate the first derivative  with respect to $Y$, and thus the higher order terms do not contribute.
The last ingredient in eq.(\ref{wf}) that requires explanation is the operator $B$. This operator is the exponential of the quadratic form of $a$'a and $a^\dagger$'s which also depends on the valence charge density $\rho_v$. It's role is to perform the Bogolyubov transformation on the soft gluon fields, mixing the creation and annihilation operators \cite{inprep}. 
The two operators in eq.(\ref{wf}), $C$ and $B$ correspond precisely to the two steps in the calculation of \cite{JIMWLK} - the expansion of the quantum fields around the classical background $b$, and subsequent integration over the soft modes keeping only quadratic terms in the Lagrangian.
The more complicated mathematics of \cite{JIMWLK} (the one loop integration over the soft modes) is coded in the operator $B$. 
However luckily for us, it has the property that for small densities it becomes a unit operator $B(\rho=0)=1$. Thus in the small density regime it only brings in perturbative corrections, and therefore in the following we set it to unity. 

In fact, once we set $B=1$ the wave function eq.(\ref{wf}) becomes precisely the one used to derive the Balitsky equation in \cite{urs}, with the substitution of the lowest order Weiszacker-Williams field $\int d^2y{x_i-y_i\over (x-y)^2}\rho^a(y)$ by $b^a_i(x)$.  Of course, $b$ which solves the nonlinear equations eq.(\ref{b}), when expanded to first order in $\rho$, reduces to this expression. 

Given the evolution of the wave function, we can now calculate the evolution of the correlation functions of the charge density. Since the boost operation  "opens up" the Hilbert space of the soft modes, we have to consider also the contribution of these modes to the charge density operator
\begin{equation}
\rho^a(x)=\rho_v^a(x)+\int^{\Lambda}_{(1-\delta Y)\Lambda} dk^+a^{\dagger b}_i(k^+,x)T^a_{bc}a^{c}_i(k^+,x)
\end{equation}
where $T^a_{bc}=if^{abc}$ is the $SU(N)$ generator in the adjoint representation. 

Sandwiching the $n$-th power of $\rho$ in the wave function eq.(\ref{wf}) and differentiating with respect to $Y$ we obtain
\begin{eqnarray}
&&{\partial\over\partial Y}\langle\rho^{a_1}(x_1)...\rho^{a_n}(x_n)\rangle={1\over \pi}
\{{1\over 2}
\int d^2x\langle[b_i^a(x),[b_i^a(x),\rho^{a_1}(x_1)...\rho^{a_n}(x_n)]\rangle+\\
&&\int d^2x\sum_{i}\delta^2(x-x_i)T^{a_i}_{bc}\langle b_i^b(x)\rho^{a_1}(x_1)...\rho^{a_{i-1}}(x_{i-1})\rho^{a_{i+1}}(x_{i+1})...\rho^{a_n}(x_n)b_i^c(x)\rangle+\nonumber\\
&&+\int d^2x\sum_{i<j}\delta^2(x-x_i)\delta^2(x-x_j)(T^{a_i}T^{a_j})_{bc}\nonumber\\
&&\times\langle b_i^b(x)\rho^{a_1}(x_1)...\rho^{a_{i-1}}(x_{i-1})
\rho^{a_{i+1}}(x_{i+1})...\rho^{a_{j-1}}(x_{j-1})\rho^{a_{j+1}}(x_{j+1})...\rho^{a_n}(x_n)b_i^c(x)\rangle+\ldots\nonumber\\
&&+\int d^2x\delta^2(x-x_i)...\delta^2(x-x_n)(T^{a_i}...T^{a_n})_{bc}\langle b_i^b(x)b_i^c(x)\rangle\}\nonumber
\label{evolwf}
\end{eqnarray}
where now the averages are over the valence state $|v\rangle$ and the charge density operators are those of the valence degrees of freedom only.

The $i$-th line in eq.(\ref{evolwf}) contains terms with $i$ operators $\rho$ deleted and substituted by the single gluon charge density operators $\delta^2(x-x_i)T^{a_i}$.
This structure is completely transparent. One step in the evolution corresponds to emission of one gluon. The first term in eq.(\ref{evolwf}) corresponds to the rotation of the valence charge density by the charge of the emitted gluon. All other terms are due to the direct contribution 
of the color charge density of the emitted gluon itself.

Note that the first (double commutator) term on the right hand side of eq.(\ref{evolwf}) despite appearances starts with the same power of $\rho$ as the left hand side. Both, in the weak and strong field limit the relevant commutator can be written as
\begin{equation}
[b^a_i(x),\rho^b(y)]={\delta b^a_i(x)\over\delta \rho^c(y)}f^{cbd}\rho^d(y)
\label{commutator}
\end{equation}
 and 
 \begin{equation}
 {\delta b^a_i(x)\over\delta \rho^c(y)}=g\left[D_i{1\over \partial\cdot D}\right]^{ac}(x,y)
 \end{equation}
 Thus each commutator reduces the lowest power of $\rho$ by one.
 This term is contained in the JIMWLK evolution at large fields, as it can be written as the contribution to the kernel with two functional derivatives with respect to $\rho$ multiplied by a function of $\rho$ whose Taylor expansion starts at $\rho^2$. It will therefore not interest us in the following, and we will concentrate our attention on the rest of the terms in eq.(\ref{evolwf}).

These terms have insertions of $T^a$ in place of $\rho^a$, and so one naturally would like to write them down as functional derivatives with respect to $\rho$. Here however one faces the problem, that the charge density operators do not commute with each other, and thus the ordering of $\rho$'s and $b$'s in eq.(\ref{evolwf}) is important. Fortunately there is a known way of representing correlation functions of noncommuting variables as a functional integral. We describe it, as well as its application to the present case in the next section.

\section{The functional integral representation and the Wess-Zumino term.}
The general problem we have to address is how to represent the correlators of the $SU(N)$ generators $\rho^a(x)$ in terms of the functional integral. We will describe this construction in the case of $SU(2)$. Let us first forget about the $x$ dependence and also assume that the generators is in the fixed representation of spin $J$.
The construction for this case has been worked out in detail in \cite{pas}.
It is is based on the observation 
that instead of considering the ordered product
 of the generators $\rho^{a}$ in the representation $J$, one can consider the correlation function
\begin{eqnarray}
\langle\rho^{a}(t_{1}) \rho^{b}(t_{2})
\ldots \rho^{c}(t_{k})\rangle \longrightarrow
J^k\langle n^{a}(t_{1}) n^{b}(t_{2})\ldots n^{c}(t_{k})\rangle =
{}~~~~~~~~~~~~~~ \\
J^k\int Dn(t) n^{a}(t_{1}) n^{b}(t_{2})\ldots n^{c}(t_{k})
\exp\left[i{J\over 2} \int_{\Sigma} d^{2}\xi
\epsilon_{\alpha\beta}\epsilon^{abc} n^{a}
\partial_{\alpha}n^{b}\partial_{\beta}n^{c}\right] \nonumber
\label{functint}
\end{eqnarray}
where $J$ is the spin of representaion (i.e. for fundamental
 representation $J = 1/2$), $n^{a}(t)$ is a unit vector
 $n^{a}n^{a} = 1$ living on a contour $C$ ($t$ is a coordinate
 on the contour) and $\Sigma$ is an arbitrary two-dimensional
 surface with the boundary $C = \delta\Sigma$.
 Despite the appearance, the
 two-dimensional Wess-Zumino action 
\begin{equation}
S[n] = \int_{\Sigma} d^{2}\xi
\epsilon_{\alpha\beta}\epsilon^{abc} n^{a}
\partial_{\alpha}n^{b}\partial_{\beta}n^{c}
\label{S[n]}
\end{equation}
 depends only on values $n^{a}(t)$
 at the boundary and not on the values $n$ takes on the surface $\Sigma$. 
 
 The coordinates $t_i$ on the left hand side of eq.({\ref{functint}) are only important as indicators of the ordering of the operators. Thus the correlation function of $n$'s must only depend on the ordering of the time coordinates, and not on their values. This is indeed the case.
 To see this note that the variation of the Wess-Zumino action under arbitrary transformation of the fields $n$ is
\begin{equation}
\delta S = \oint_{C} dt \epsilon^{abc} n^{a}
\partial_{t}n^{b} \delta n^{c}
\label{Svariation}
\end{equation}
Now let us perform an infinitesimal $SU(2)$ transformation 
\begin{equation}
n^{a}(t) \rightarrow n^{a}(t) + \epsilon^{abc} \Omega^{b}(t)
n^{c}(t)
\end{equation}
The action change under this transformation is $\delta S = -
\oint_{C} dt \dot{n}^{a}(t)\Omega^{a}(t)$ 
Performing this transformation in the path integral for the correlation function eq.(\ref{functint}) we obtain
\begin{eqnarray}
J\frac{d}{dt}\langle n^{a}(t) n^{b}(t_{1})\ldots n^{c}(t_{k})\rangle= i
\sum_{i=1}^{k}\delta(t-t_{i})\epsilon^{adf}\langle n^{f}(t)
n^{b}(t_{1}) \ldots \underline{n^{d}(t_{i})}
\ldots  n^{c}(t_{k})\rangle
\label{ward}
\end{eqnarray}
where $\underline{n^{d}(t_{i})}$ means the exclusion of this term
 from the products of the fields in a correlator. This establishes the piecewise constant nature of the correlation function. Moreover, remembering that in the equal time limit the time derivative of the correlation function reduces to the equal time commutator (as the path integral represents $T$-ordered products),
one concludes immediately from eq.(\ref{ward}) that the following equal time commutaion relations hold
\begin{equation}
J\left[n^{a}, n^{b}\right] = i \epsilon^{abc} n^{c}
\end{equation}
 This establishes that the following identification is valid
 \begin{equation}
 \rho^a=Jn^a
 \end{equation}

 The fact that the charge density depends on the transverse coordinate simply means that we have to make the unit field $n^a$ also $x$-dependent, but the Wess-Zumino term is strictly local in $x$ (since otherwise we would introduce noncommutativity between $\rho$'s at different points in the transverse plain). Finally we should also allow to consider states with different representation of the $SU(2)$ group. This is achieved by allowing $J$ to be distributed with some arbitrary weight and supplementing the integration measure over $n^a$ by summation over all half integer $J$.
 
 All said and done we see that we can represent the calculation of charge density correlators by the following functional integral
 \begin{equation}
 \langle\rho^{a_1}(x_1)...\rho^{a_n}(x_n)\rangle=\int d\rho(x,t)\rho^{a_1}(x_1,t_1)...\rho^{a_n}(x_n,t_n)
 W[\rho]
 \label{tr}
 \end{equation}
 with the measure of integration being understood as
 \begin{equation}
 \int d\rho(x,t)\ldots=\int\Sigma_{J(x)} \Pi_{x,t} dn(x,t)\exp\left[i\int d^2x {J(x)\over 2} \int_{\Sigma} d^{2}\xi
\epsilon_{\alpha\beta}\epsilon^{abc} n^{a}(x,t)
\partial_{\alpha}n^{b}(x,t)\partial_{\beta}n^{c}(x,t)\right]\ldots
\label{measure}
\end{equation}

The generalization of this procedure to the $SU(N)$ group is somewhat cumbersome and we will not describe it in detail, but rather only indicate the flow of the argument. It goes along the following lines. The Wess-Zumino term in eq.(\ref{functint}) can be thought of as the flux of the magnetic field of the 'tHooft-Polyakov monopole through the contour $\Sigma$. This monopole would sit in some three dimenisonal space, two of whose dimensions are spanned by the coordinates on the surface $\Sigma$. It is of course purely fictitious and should not be given any physical significance beoynd a simple mnemonic to describe the mathematical structure of the relevant Wess-Zumino term. The conservation of the magnetic flux of this monopole modulo $4\pi$ is the reason why the Wess-Zumina term does not depend on the surface $\Sigma$, but rather only on its boundary, as long as the coefficient $J$  takes half integer values. In the $SU(N)$ group there are $N-1$ such independent magnetic monopoles, whose magnetic fields and therefore fluxes can be expressed in terms of the adjoint "Higgs field" of unit length $n^a$, $a=1,...,n^2-1$\cite{monop}. 
Thus there are $N-1$ independent Wess-Zumino terms that can be written for the $SU(N)$ group. The coefficients of all these terms have to be quantized for the same reason as $J$, and this quantization corresponds to the quantization of the eigenvalues of the generators
of the Cartan subalgebra of the $SU(N)$ group. To handle the generators of $SU(N)$ in a given representation one again has to endow $\rho^a$ with a dependence on one extra coordinate $t$, and introduce $N-1$ Wess-Zumino terms in the integration measure. The coefficients of these terms should be equal to the values of the Cartan generators in the highest weight state of the given representation. Taking into account $x$ dependence and allowing for the variation of the representation is achieved in the same way as in eq.(\ref{measure}).

We note that in the limit of large representations, $J\rightarrow\infty$, the  Wess-Zumino term imposes the constraint $\dot n^a(t)=0$. Thus the $t$-dependence of the unit vector $n$ is frozen, $n^a(t)=n^a$, so that one recovers the formulation of  eq.(\ref{average}).

Practically speaking therefore we see that taking into account noncommutativity of $\rho$ leads to two major changes, first the field in the functional integral becomes dependent on extra coordinate $t$, and second the measure of the integration becomes complex with the phase given by the Wess - Zumino term. 

Nevertheless, given the functional integral representation, we can conveniently rewrite the evolution equations eq.(\ref{evolwf}) as a functional derivative operator acting on $W[\rho]$.
It is in fact easy to see that the whole hierarchy of eqs.(\ref{evolwf}) can be written in a very coincise form
\begin{eqnarray}
&&\int d^2x\sum_{i}\delta^2(x-x_i)T^{a_i}_{bc}\langle b_i^b(x)\rho^{a_1}(x_1)...\rho^{a_{i-1}}(x_{i-1})\rho^{a_{i+1}}(x_{i+1})...\rho^{a_n}(x_n)b_i^c(x)\rangle=\\
&&\langle\int d^2xb_i^b(x,t\rightarrow-\infty)T^{a}_{bc}b_i^c(x,t\rightarrow+\infty)\int dt {\delta\over\delta\rho^a(x,t)}\{\rho^{a_1}(x_1,t_1)...\rho^{a_n}(x_n,t_n)\}\rangle;\nonumber\\
&&\int d^2x\sum_{i<j}\delta^2(x-x_i)\delta^2(x-x_j)(T^{a_i}T^{a_j})_{bc}\nonumber\\
&&\times\langle b_i^b(x)\rho^{a_1}(x_1)...\rho^{a_{i-1}}(x_{i-1})\rho^{a_{i+1}}(x_{i+1})...\rho^{a_{j-1}}(x_{j-1})\rho^{a_{j+1}}(x_{j+1})...\rho^{a_n}(x_n)b_i^c(x)\rangle=\nonumber\\
&&\langle\int d^2xb_i^b(x,t\rightarrow-\infty)(T^{a}T^d)_{bc}b_i^c(x,t\rightarrow+\infty)\int_{-\infty}^{t'} dt\int_{-\infty}^{+\infty}dt' {\delta\over\delta\rho^a(x,t)}{\delta\over\delta\rho^b(x,t')}\{\rho^{a_1}(x_1,t_1)...\rho^{a_n}(x_n,t_n)\}\rangle\nonumber\\
&&\ \ \ \ \ \  \ etc...\nonumber
\end{eqnarray}

Thus somewhat surprisingly we can rewrite the right hand side of eq.(\ref{evolwf}) (subtracting the first term) as
\begin{equation}{1\over \pi}
\langle\int d^2xb_i^b(x,t\rightarrow-\infty)\left[P\exp\left\{\int_{-\infty}^{+\infty} dt T^{a}{\delta\over\delta\rho^a(x,t)}\right\}-1\right]_{bc}b_i^c(x,t\rightarrow+\infty)\{\rho^{a_1}(x_1,t_1)...\rho^{a_n}(x_n,t_n)\}\rangle
\label{pathint}
\end{equation}
where $P$ denotes the path ordering along  $t$. 
Note that the linear and quadratic terms in the expansion of the path ordered exponential are contained already in the JIMWLK equation (modulo the noncommutativity of $\rho$'s). The rest of the terms are subleading at large $\rho$. However at $\rho=O(1)$ all the terms in eq.(\ref{pathint}) are of the same order and should be kept. Also note, that by keeping $b^a_i$ as the full solution of the equation eq.{\ref{b}) and not its leading perturbative term, we are resumming some terms that are subleading in the low density limit.
Partially integrating the functional derivatives in the functional integral representation eq.(\ref{tr}) we can finally rewrite the evolution as the functional equation for $W$
\begin{eqnarray}
&&{\partial\over\partial Y}W[\rho(x,t)]=\alpha_s\left\{O_{JIMWLK}(\rho,{\delta\over\delta\rho})W[\rho(x,t)]\right\}\nonumber\\&&+{1\over \pi}\left\{
\int d^2xb_i^b(x,t\rightarrow-\infty)\left[\tilde P\exp\left\{-\int_{-\infty}^{+\infty} dt T^{a}{\delta\over\delta\rho^a(x,t)}\right\}-1\right]_{bc}b_i^c(x,t\rightarrow+\infty)\right\}W[\rho(x,t)]
\label{final}
\end{eqnarray}
where $\tilde P$ is the path ordered exponential with the linear and quadratic terms subtracted.

This is the main result of the present paper.

\section{Discussion.}

There are several interesting questions that arise from the previous derivation. First, does the variable $t$ has a physical meaning beoynd being a useful tool to represent correlators of noncommuting variables. We believe that the answer to this question is in the affirmative.
We remind the reader that one already had a need in introducing path ordering in the discussion of high energy evolution. In particular it was realized in \cite{jkmw} that this was necessary in order to solve the classical equations of motion for the gluon field in the presence of the charge density $\rho$. The equation considered in \cite{jkmw} was identical to eq.(\ref{b}), except $b$ was considered to be a classical field which depended on $x$ as well as on the longitudinal coordinate $x^-$. It then turned out that only one contribution to the commutator term in the first equation of eq.(\ref{b}) was important. This contribution was the one where one of the $b$'s had the longitudinal coordinate slightly smaller than that of the other $b$. The equation then was formally solved as
\begin{equation}
b_i(x,x^-)=iU^\dagger(x,x^-)\partial_iU(x,x^-), \ \ \ \ \ \ \ U(x,x^-)=P\exp\{i\int_{-\infty}^{x^-}dx^-\alpha(x,x^-)\}
\end{equation}
with the function $\alpha$ itself defined in terms of the matrix $U$ and the charge density $\rho$ as $\alpha={g^2\over\partial^2}[U\rho U^\dagger]$. The important element in these expressions is the path ordering with respect to $x^-$. 
Now returning to our present framework we can ask ourselves what is the solution of the operator equation eq.(\ref{b}) that appears in the functional integral expression eq.(\ref{final}). It is obvious, that in order to solve this equation one has to follow the procedure identical to that in \cite{jkmw}. One has to endow $b$ and $\rho$ with the extra coordinate $t$, the commutator term then has the $t$ coordinate of the first $b$-factor slightly smaller than that of the second factor, as the $t$-ordering in the classical equation simply reflects the 
operator ordering in the operator equation. From that point on the solution is identical to that of \cite{jkmw} with the path ordering in $x^-$ replaced by the path ordering in $t$.

We conclude therefore that the ordering variable $t$ that we have introduced in this paper is identical to the longitudinal coordinate $x^-$\cite{foot}.

One than immediately is tempted to ask: what about the other variable which is introduced to define the Wess-Zumino term? If $t$ is $x^-$, then the other one may be $x^+$.
 The connection here is more tenuous, but we believe it is true. The Wess-Zumino term can be thought of as the Berry phase \cite{berry} for the state in the J-representation of the charge density operator. The state depends on $x^-$ as a parameter, and its phase changes as this parameter changes adiabatically along the hadron. The Berry phase arises as a topological part of the time integral of the simplectic form in the action. The QCD functional integral in the light cone gauge (to which the functional integral eq.(\ref{functint}) is supposed to be an approximation)  is in fact a phase space path integral, and contains the simplectic form $F^{+i}F^{-i}=\partial^+b_i\partial^-b_i$. It is likely that the Wess-Zumino term in eq.(\ref{functint}) should be properly understood as the time integral of this simplectic form, and thus the second coordinate which appears in its definition is indeed the time variable $x^+$. This is an interesting question, which should be understood better.
 
Another natural question that arises, is whether one can somehow avoid introducing the noncommuting variables alltogether, and thus get rid of the awkward phase in the "weight functional" $W$. Unfortunately we do not believe this possible. Suppose we are interested in the expectation values of some set of observables $\{O_i\}$. If all these observables are mutually commuting, one can choose a basis in the Hilbert space (spanned by a set of coordinates which commute with  $\{O_i\}$) such that all the expectation values are given as averages over the real measure. The measure is the square of the wave function in the basis we have chosen. However if we are also interested in averages of other variables which do not commute with $O_i$, the calculation of the expectation value will necessarily probe also the phase of the wave function.
The question is thus whether in high energy QCD we are only interested in mutually commuting observables. The observables we would like to calculate in the hadronic wave function are the averages of the $S$-matrix of fast particles scattering on it. In general these observables do not commute. Physically it simply means that the probability for scattering of two particles depends on the order in which these particles scatter, since the first one to scatter perturbs the target fields. When the number of the projectile particles is small and the target fields are large, this small perturbation is a subleading effect and thus can be neglected. However for targets which are not dense this noncommutativity is important, and so the phase of the wave function comes into play.

Perhaps the most interesting question that is suggested by our result eq.(\ref{final}) is whether there is some well defined duality transformation that maps the high density regime into the low density regime. Recall, that in the high density JIMWLK expression the prominent element is the matrix $U$ defined as 
\begin{equation}
U=P\exp\{i \int dx^-\alpha(x,x^-)\}
\end{equation}
The relation between $\alpha$ and $\rho$ is fairly complicated, but at low density it simplifies considerably
\begin{equation}
\alpha(x,x^-)=_{low\ \ density}{g^2\over\partial^2}(x,y)\rho(y,x^-)
\label{alpharho}
\end{equation}
Thus it looks like the path ordered exponential of density and the path ordered exponential of functional derivatives with respect to density roughly exchange their roles in the high and low density regimes. Of course the swap is not so starightforward, but nevertheless the appearance of the path order exponential of $\delta\over\delta\rho$ in eq.(\ref{final}) is very suggestive that such a transformation can be defined. One could hope that the knowledge of this transformation will help find the exact expression for the operator of the high energy evolution valid at arbitrary density $\rho$. 

Finally we comment on the relation of our results to those presented in \cite{shoshi}. The paper \cite{shoshi} is not explicit about the operator ordering of the operators involved, and that makes the detailed comparison a little ambiguous. Also, the derivation in \cite{shoshi} are only given in the large $N_c$ approximation which relies on the dipole model. However it is easy to see that the general structure of their result is the same as our expression eq.(\ref{final}) expanded to fourth order in the functional derivative. To establish this one has to use the perturbative relation between $\alpha$ and $\rho$ eq.(\ref{alpharho}) and the commutator of the field $b$ with $\rho$ eq.(\ref{commutator}) taken to lowest order in the strong coupling constant.  Additionally one has to assume that the funtional $W$ depends only on the "dipole cross section" $s(x,y)={1\over N}{\rm tr} [U^\dagger(x)U(y)]$.
\acknowledgments
We thank Urs Wiedemann for many discussions on subjects directly relevant to this work.


\begin{thebibliography}{99}
\bibitem{foot} To match the dimensions of $t$ and $x^-$ one should scale $x^-$ by the longitudinal extent of the target.
\end{thebibliography}
\end{document}